\documentclass[
aps,
pra,
twocolumn,
amsmath,
amssymb,
nofootinbib,
showpacs,
superscriptaddress,
reprint
]{revtex4-2}

\usepackage{latexsym}
\usepackage{graphics}
\usepackage{graphicx}
\usepackage{epsfig}
\usepackage{color}
\usepackage{amsmath}
\usepackage{amssymb}
\usepackage{amsthm}
\usepackage{dcolumn}
\usepackage{bm}
\usepackage{float}
\usepackage{hyperref}
\usepackage{color}
\usepackage{epstopdf}
\usepackage{braket}
\usepackage{cleveref}
\usepackage[svgnames]{xcolor}
\hypersetup{hidelinks,colorlinks=true,allcolors=DarkBlue}
\usepackage[mathlines]{lineno}
\usepackage{soul} 

\usepackage[export]{adjustbox}

\newcommand{\comment}[1]{}

\DeclareMathOperator{\Tr}{Tr}
\theoremstyle{remark}

\begin{document}
\preprint{APS/123-QED}

\title{Eurasian-Scale Experimental Satellite-based Quantum Key Distribution \\ with Detector Efficiency Mismatch Analysis}

\author{Aleksandr V. Khmelev}
    \email{aleksandr.khmelev@phystech.edu}
    \affiliation{Russian Quantum Center, Skolkovo, Moscow 121205, Russia}
    \affiliation{Moscow Institute of Physics and Technology, Dolgoprudny, Moscow Region 141700, Russia}
    \affiliation{QSpace Technologies, Moscow 121205, Russia}
\author{Alexey V. Duplinsky}
    \affiliation{QSpace Technologies, Moscow 121205, Russia}
    \affiliation{HSE University, Moscow 101000, Russia}
\author{Ruslan M. Bakhshaliev} 
    \affiliation{QSpace Technologies, Moscow 121205, Russia}
\author{Egor I. Ivchenko}
    \affiliation{Russian Quantum Center, Skolkovo, Moscow 121205, Russia}
    \affiliation{Moscow Institute of Physics and Technology, Dolgoprudny, Moscow Region 141700, Russia}
    \affiliation{QSpace Technologies, Moscow 121205, Russia}
    \affiliation{National University of Science and Technology MISIS, Moscow 119049, Russia}
\author{Liubov V. Pismeniuk} 
    \affiliation{QSpace Technologies, Moscow 121205, Russia}
\author{Vladimir F. Mayboroda}
    \affiliation{National University of Science and Technology MISIS, Moscow 119049, Russia}
\author{Ivan S. Nesterov}
    \affiliation{Russian Quantum Center, Skolkovo, Moscow 121205, Russia}
    \affiliation{QSpace Technologies, Moscow 121205, Russia}
\author{Arkadiy N. Chernov}
    \affiliation{Russian Quantum Center, Skolkovo, Moscow 121205, Russia}
    \affiliation{Moscow Institute of Physics and Technology, Dolgoprudny, Moscow Region 141700, Russia}
    \affiliation{QSpace Technologies, Moscow 121205, Russia}
    \affiliation{National University of Science and Technology MISIS, Moscow 119049, Russia} 
\author{Anton S. Trushechkin}
    \affiliation{National University of Science and Technology MISIS, Moscow 119049, Russia}
    \affiliation{Steklov Mathematical Institute of Russian Academy of Sciences, Moscow 119991, Russia}
\author{Evgeniy O. Kiktenko}
    \affiliation{Russian Quantum Center, Skolkovo, Moscow 121205, Russia}
    \affiliation{National University of Science and Technology MISIS, Moscow 119049, Russia}
\author{Vladimir L. Kurochkin}
    \email{v.kurochkin@rqc.ru}
    \affiliation{Russian Quantum Center, Skolkovo, Moscow 121205, Russia}
    \affiliation{Moscow Institute of Physics and Technology, Dolgoprudny, Moscow Region 141700, Russia}
    \affiliation{QSpace Technologies, Moscow 121205, Russia}
    \affiliation{National University of Science and Technology MISIS, Moscow 119049, Russia}
\author{Aleksey K. Fedorov}
    \email{akf@rqc.ru}
    \affiliation{Russian Quantum Center, Skolkovo, Moscow 121205, Russia}
    \affiliation{National University of Science and Technology MISIS, Moscow 119049, Russia}

\date{\today}
\begin{abstract}
The Micius satellite is the pioneering initiative to demonstrate quantum teleportation, entanglement distribution, quantum key distribution (QKD), and quantum-secured communications experiments at the global scale. In this work, we report on the results of the 600-mm-aperture ground station design which has enabled the establishment of a quantum-secured link between the Zvenigorod and Nanshan ground stations using the Micius satellite. As a result of a quantum communications session, an overall sifted key of 2.5 Mbits and a total final key length of 310 kbits have been obtained. We present an extension of the security analysis of the realization of satellite-based QKD decoy-state protocol by taking into account the effect of the detection-efficiency mismatch for four detectors. We also simulate the QKD protocol for the satellite passage and by that validate our semi-empirical model for a realistic receiver, which is in good agreement with the experimental data. Our results pave the way to the considerations of realistic imperfection of the QKD systems, which are important in the context of their practical security.
\end{abstract}

\maketitle

\section{Introduction}\label{sec:level1}

The development of computational devices opens new possibilities to attack cryptographic systems, whose security is based on computational assumptions. 
In particular, powerful enough quantum computing devices have the potential to compromise the security of widely deployed cryptographic tools~\cite{Shor, Grover}, such as those based on the complexity of integer factorization~\cite{RSA}.
One of the solutions is to switch to information security tools~\cite{Gisin2002,Scarani2009,Diamanti2016,Bernstein2017,Yunakovsky2021}, 
which are considered to be resistant to attacks with quantum computers. 
Specifically, quantum key distribution (QKD), which is the technology that allows establishing secure communications between distinct parties with the security guaranteed by the law of physics~\cite{Gisin2002,Scarani2009}, offers such a possibility.  
The idea behind QKD is to encode information in states of single photons and exchange them in a way that prevents uncontrollable eavesdropping.
By following such a QKD protocol properly, two distant parties share strings of random bits --- quantum-generated secret keys.
Although remarkable progress in QKD has been performed~\cite{Gisin2002,Scarani2009,Diamanti2016}, 
one of the most important problems is photon loss in the channel, 
so that the distance of peer-to-peer QKD with reasonable key rates is about thousand of kilometers~\cite{Pan2021, Liu2017}.
This seems to be a limitation to applying the QKD technology at the global scale. 

Overcoming this challenge of the distance is possible with the use of satellite-to-ground QKD~\cite{Pan2017,Pan2018,Pan20182,Pan20212}.
The launch of a low-Earth-orbit Micius satellite for implementing decoy-state QKD~\cite{Pan2017},
realizing intercontinental quantum-secured communications~\cite{Pan20182}, 
and deploying an integrated space-to-ground quantum communication network of the length over 4,600 kilometres~\cite{Pan20212}, 
have been successfully demonstrated during the last years (for a review, see Refs.~\cite{Bedington2017,Pan2022}). 
The overall losses for the satellite-to-ground link are small compared to ground-level transmission due to negligible attenuation 
above the atmosphere and insignificant vertical turbulence in the lower atmosphere.
\begin{figure*}[ht!]
\centerline{\includegraphics[width=1\linewidth]{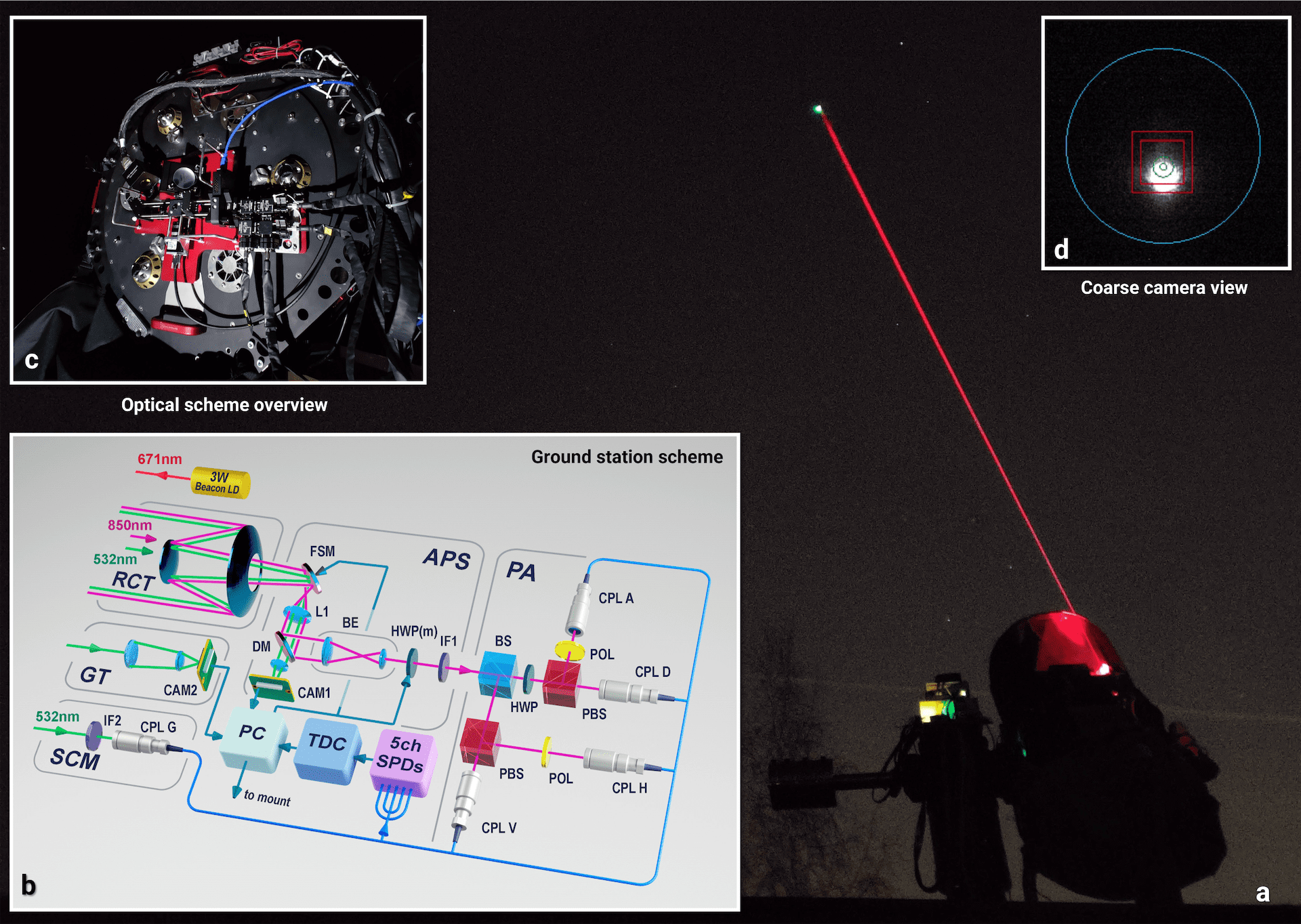}}
\caption{Zvenigorod ground station overview. 
(\textbf{a}) The general view of the satellite-to-ground quantum communication experiment from the Zvenigorod ground station. 
The satellite sends 532\,nm laser pulses to the ground, and the receiver returns the 671\,nm laser beam. 
As a result, a stable satellite-to-ground link is established.
(\textbf{b}) The schematic of the ground station. 
The quantum (850\,nm) and beacon (532\,nm) laser signals are received by the Ritchey-Chretien telescope (RCT), 
and their directions are regulated by the fast steering mirror (FSM) depending on the location of the beacon beam on the fine camera (CAM1). 
To decode quantum states, the classical passive BB84 scheme or polarization analyzer 
(PA: BS is the 50:50 beam splitter; PBS is the polarizing beam splitter; HWP is the half-wave plate; 
POL is the polarizer) is used after spectrum filtering (DM is the dichroic mirror; IF1 is the interference filter) 
and polarization compensation~\cite{duplinsky} (HWP-m is the motorized half-wave plate). 
Then, a single photon detector module (5 ch SPDs) and a time-to-digital converter (TDC) 
register the light gathered by the four couplers (CPL {H,V,D,A}) 
and beacon pulses collected by the synchronization module (SCM: IF2 is the interference filter; CPL G is the coupler).
To point the telescope at the satellite, we use a guide telescope (GT) with a coarse camera (CAM2).
(\textbf{c}) Hardware of the optical part at Zvenigorod ground station. 
(\textbf{d}) The coarse camera view during the ground station tracking of the Micius satellite. 
The white spot on the frame is the satellite, and the green circle is the target position.
\label{scheme}}
\end{figure*}
Several satellite-based QKD research projects demonstrating diverse technology concepts have been initiated~\cite{sidhu2021advances, QEYSSat, Toyoshima, Takenaka, Bourgoin, Villar, Anwar}. 
The various QKD models~\cite{VERGOOSSEN, Sidhu, ecker, MDIModel,Liorni,Abasifard} investigating feasible concepts and designs of quantum networks in space 
have been significantly developed since the first satellite-based quantum communication experiments. 
However, these models commonly use a limited dataset from the quantum experiments provided by the Micius satellite to validate these QKD models. 
Hence, it results in restricted validation of satellite-based QKD models. 
Specifically, they do not take into account several potentially important imperfections in the realization of QKD protocols. 
For example, the problem of detector efficiency mismatch~\cite{Bochkov, Trushechkin, Ivchenko}, 
which has been recently investigated in the context of the decoy-state QKD protocol~\cite{LoMa,Trushechkin2021},
seems to be important for consideration in the security analysis.

In this paper, 
we report on the results of establishing a quantum-secured link between the Zvenigorod and Nanshan ground stations using the Micius satellite. 
We realize the decoy-state QKD protocol and obtain a sifted key length of 2491~kbits and a total final key length of 310~kbits.
This has enabled transferring messages that are encrypted via a one-time pad with quantum-generated keys over 3800 km. 
Also, we present a modification of the security analysis of the realization of satellite-based QKD decoy-state protocol by taking into account the effect of the detection-efficiency mismatch. 
We also simulate the QKD protocol for that satellite passage, validate our semi-empirical model for a realistic receiver, and find its agreement with the experimental data. 
Our results pave the way to the considerations of realistic imperfection of the QKD systems, 
which are important in the context of their practical security.

The paper is organized as follows. 
In Sec.~\ref{sec2}, we describe the developed ground station and report the results of the joint QKD experiment between 
the Micius satellite and the Zvenigorod ground station. 
In Sec.~\ref{sec3}, we provide a semi-empirical model of satellite-to-ground QKD and compare its outcome to experimental results. 
The four detector efficiency mismatch analyses of secure key generation 
during satellite-to-ground quantum communication are considered in Sec.~\ref{sec4}.
Finally, in Sec.~\ref{sec5}, we provide the results of the quantum secure message transfer between Nanshan and Zvenigorod; 
we summarize our results and conclude in Sec.~\ref{sec5}

\section{Satellite-to-ground QKD experiment}\label{sec2}

In this section, we describe our ground station~\cite{TechLett} that is equipped with a 600-mm-aperture reflector telescope and placed at the Zvenigorod observatory 
(Moscow region, $55$\textdegree~$42^\prime\,$N, $36$\textdegree~$45^\prime\,$E, altitude of 198~m). 
The main characteristics are listed in Tables~\ref{TabS}, \ref{tab1}, and \ref{tabA1}. 
Also, we present the results of the joint satellite-to-ground QKD experiment and provide the observable optical characteristics and noise level. 
A general scheme of the quantum communication procedure is given in Fig.~\ref{scheme}a.

\subsection{Zvenigorod ground station}\label{sec2.1}

The ground station has a two-stage acquiring, pointing, and tracking (APT) system that ensures high angular stability for the optical signal 
and suitable operating conditions for the photon states decoder during quantum communication. 
The general scheme and the hardware at the Zvenigorod ground station are illustrated in Fig.~\ref{scheme}b and Fig.~\ref{scheme}c, respectively. 

The coarse control of the station is achieved by a motorized mount of the primary telescope, 
which uses the calculated satellite location in orbit, and an auxiliary \mbox{70-mm} telescope with a camera for pointing to the satellite. 
The algorithm on the camera determines the image center of the satellite, and the telescope orientation is adjusted. 
Figure~\ref{scheme}d shows a frame from the coarse camera during satellite tracking. 
The tracking error of the coarse control stage is typically less than 100\,$\mu$rad.

\begin{figure*}[t!]
\includegraphics[width=0.93\linewidth]{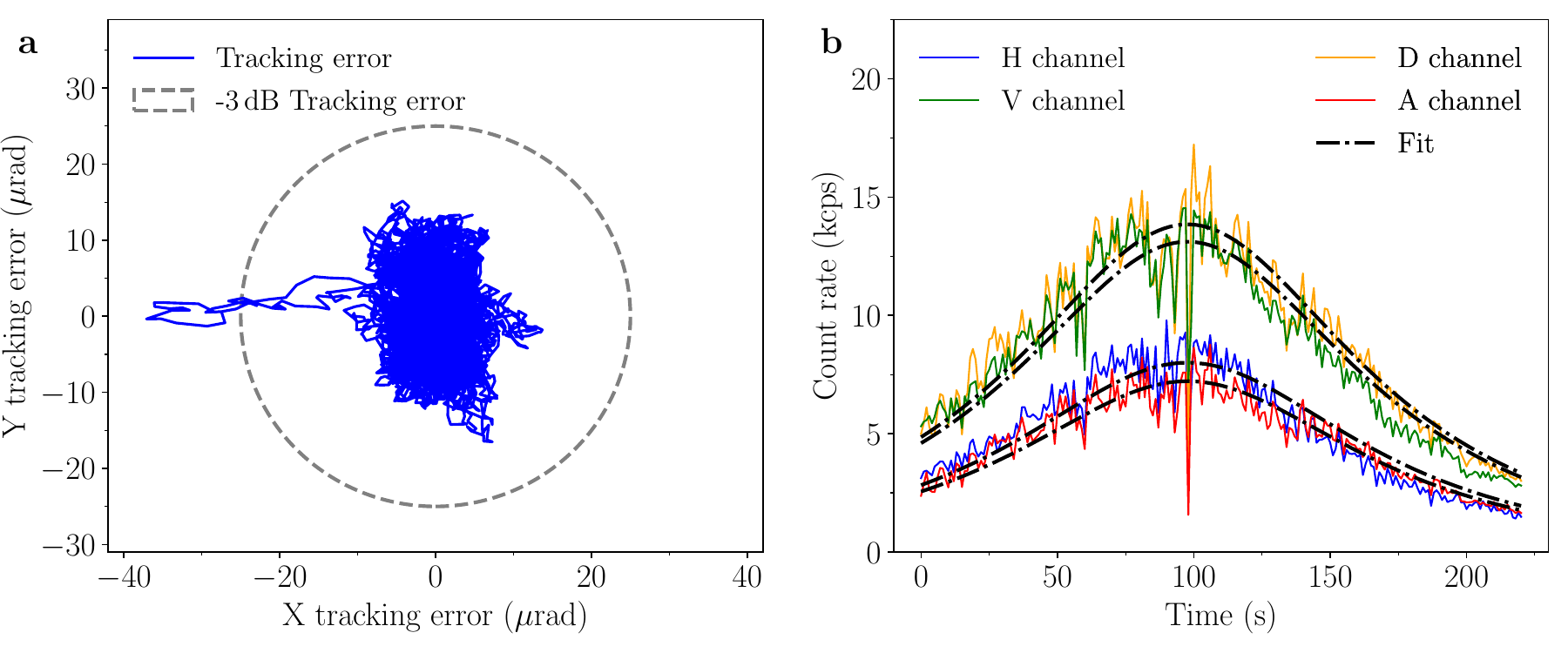}
\caption{ (\textbf{a})\,The tracking error of the X and Y axes obtained from real-time frames read out from the fast camera (CAM1). 
The circle illustrates the tracking inaccuracy when the quantum signal drops by\,3\,dB. 
(\textbf{b})\,The photon count rate for four channels during the satellite-to-ground QKD experiment. 
The dash-dotted lines represent the function $\mathcal{F}_\xi (t)$ fitted the data of each $\xi$-type channel of the ground receiver.
\label{Err&RawKey}}
\end{figure*} 

The ground station features an optical signal analysis and processing system (APS) for adjusting the beam deflection, 
rotating the polarization reference frame~\cite{duplinsky}, and spectrum filtering the photons. 
The main task of APS is the fine control of the beams, which includes a fast-steering mirror and a high-speed camera. 
As a result, the high-frequency deflections of the co-aligned beacon 
beam with quantum signals are corrected to an accuracy of less than 10\,$\mu$rad. 
Moreover, there are spectrum filters to suppress the background noise and a motorized half-wave plate to reduce decoding errors (see Sec.~\ref{sec3}).

The polarization analyzer is configured in the classical BB84 scheme with a 50:50 beam splitter 
and includes four types of channels: horizontal (H), vertical (V), diagonal (D), and antidiagonal (A) polarization.
To distinguish orthogonal quantum states, two polarization beam splitters with additional polarizers on the reflection side are utilized. 
The mean polarization contrast ratio of the decoder is more than 350:1, which is measured in multiphoton mode. 
The quantum signals are coupled into four optical fibers with a core diameter of 105\,$\mu$m and detected by the single-photon module. 
The time-to-digital converter with picosecond resolution is used to record the registration moment.

Synchronization pulses are also detected by the single-photon module, and detection moments are recorded by the time-to-digital converter. 
Such an extra light channel allows us to find a correlation between the quantum states transmitted from the satellite 
and recorded on the ground station for the following sifting procedure~\cite{AperiodicSync,QubitSync}, 
as well as to suppress the background noise with temporal filtering~\cite{Pan2017}. 


\subsection{Joint QKD experiment with the Micius satellite}\label{sec2.2}

The joint QKD experiment between the Micius satellite and the Zvenigorod ground station took place on March 1, 2022,
at night with low background noise. We received quantum states from the Micius satellite for roughly 220 seconds,
with a reception angle area (-3\,dB) for the four co-aligned channels of 50\,$\mu$rad. 
\mbox{Figure~\ref{Err&RawKey}a} shows the tracking error of the satellite beacon laser, 
which was typically less than 10\,$\mu$rad and corresponded to the angle deviations of the quantum signal.

The Micius satellite passage had a peak elevation angle of 54\textdegree~and a maximum photon count rate was more than 49 kHz (see Fig.~\ref{Err&RawKey}b), while the distance between the satellite and the ground station varied from 600\,km to 1,100\,km. 
The stable bidirectional tracking started at an elevation angle of 28\textdegree. 
The operating elevation \mbox{angles} of the Zvenigorod ground station ranges from 20\textdegree~to 90\textdegree~above the horizon.

The measured photon count rates differ for the four states, 
which is mostly due to auxiliary polarization filtering on the reflection sides of PBS (see Fig.~\ref{scheme}b) that affects their optical efficiency.
To determine the observable transmission efficiency of the $\xi$-type channel, 
we assume that the photon count rate can be described as follows:
\begin{equation}\label{fit}
\mathcal{F}_\xi (t)=  p_\xi [ (T\eta_{\xi}(t) +C) +  \sum\limits_{\alpha} f p_{\alpha} (1 - e^{-\alpha\eta_{\xi}(t)})]~, 
\end{equation}
where $p_\xi$ is the fraction of photons among all received ones that enter the $\xi$-type channel 
(for the symmetric BB84 protocol with equal probability of basis choice for both transmitter and receiver 
$p_\xi = 1/4$, $\forall\,\xi\,\in\,\{\rm H,V,D,A\}$), 
$\eta_\xi(t)$ is the time-dependent overall transmission efficiency of the $\xi$-type channel at 850 nm between the Micius satellite 
and the ground station, $f$ is the repetition frequency of the quantum signals, $\alpha\,\in\,\{\mu, \nu,\lambda\}$ are the intensities 
of signal $\mu$, decoy $\nu$, and vacuum $\lambda$=0 coherent states and $p_{\alpha}\in\{p_s, p_d, p_v\}$ 
are their probabilities to be sent by satellite, respectively. 
The fixed coefficients $T$ and $C$ in Eq.~(\ref{fit}) denote satellite noise and constant background noise, respectively. 
Appendix~\ref{appendix:B} specifies the coefficients in detail. 
The next components of the function $\mathcal{F}_\xi (t)$ correspond to the count rate for a source with the intensity of 
signal, decoy, and vacuum states, according to the general model~\cite{LoMa,Trushechkin2021}. 

The overall link efficiencies $\eta_\xi(t)$ of the $\xi$-type channel according to our semi-empirical 
satellite-to-ground QKD model~\cite{SemiEmpiModel} are given by:
\begin{equation}\label{link}
\eta_\xi(t) = \frac{\varepsilon D_T^2}{(\gamma d)^2} \cdot 10^{
-0.4\varkappa \csc{\theta_{\rm El}} \cdot (1-0.0012 \cot^2{\theta_{\rm El}})
} \eta_{\mathrm{opt,}\xi} \eta_{\rm det}~,
\end{equation}
where $D_T$ is the telescope diameter, 
$\varepsilon$ is the obstruction of the telescope's secondary mirror, 
$\gamma$ is the laser source divergence,
$\eta_{\mathrm{opt,}\xi}$ is the optical efficiency of the $\xi$-type channel, 
$\eta_{\rm det}$ is the quantum efficiency of the single-photon detectors. 
TLE data is used to calculate the communication channel length $d=d(t)$ 
and satellite elevation angle $\theta_{\rm El}=\theta_{\rm El}(t)$ above the horizon, which vary with time.
Table~\ref{TabS} and Table~\ref{tabA1} provides the satellite photon source and the receiver parameters, respectively.

\begin{table}[h]
\caption{\label{TabS}Key parameters of the Micius quantum source~\cite{Pan2017}.}
\begin{ruledtabular}
\begin{tabular}{ccccccc}
$\mu$\ & $\nu$ & $p_{s}$ & $p_{d}$ & $p_{v}$ & $f\,{\rm (Hz)}$ & $\gamma\,{\rm (Rad)}$
\\ \colrule
 0.8 & 0.1 & 0.5 & 0.25 & 0.25 & $10^8$ & $10^{-5}$\\
\end{tabular}
\end{ruledtabular}
\end{table}

Hence, we determine two variables in Eqs.~(\ref{fit}) and~(\ref{link}), such as the atmospheric extinction coefficient $\varkappa = 0.22\pm0.04$, 
which corresponds to clear weather conditions, and the optical efficiency for each $\xi$-type channel, which is
specified in Table~\ref{tab1}. 
\begin{table}[H]
\caption{Optical efficiencies of the $\xi$-type channels} \label{tab1}
\begin{ruledtabular}
\begin{tabular}{ccccc} 
	& \textbf{H\,channel} & \textbf{V\,channel} & \textbf{D\,channel} & \textbf{A\,channel} 
    \\
    \colrule 
    $\eta_{\mathrm{opt,}\xi}, \%$ & $21\pm1$ & $35\pm1$ & $37\pm1$ & $19\pm1$  
\end{tabular}
\end{ruledtabular}
\end{table}

To get the sifted key, we achieve clock synchronization~\cite{AperiodicSync,QubitSync} 
between the satellite and the ground station with a precision ($\sigma$) of 500\,ps (see Fig.~\ref{sync}). 
Then the temporal filtration with a 2\,ns time window is performed to increase the signal-to-noise ratio.

\section{Experiment and model validation}\label{sec3}

For the characterization of the experimental results obtained in the joint QKD experiment 
on March 1, 2022, the simulation of the source of quantum states, 
time-dependent communication channels and optical characteristics of the receiver is completed.
Here we consider a modified semi-empirical satellite-to-ground QKD model~\cite{SemiEmpiModel} 
implied for the Zvenigorod ground station and the Micius satellite to determine the bit generation rate after basis reconciliation 
and quantum bit error rate (QBER) for the sources of the signal, decoy, and vacuum states.

Taking into account the variance in optical efficiency among the $\xi$-type channels for the ground receiver, 
we adapt the calculation of the bit generation rate (see Eq.~(9) in Ref.~\cite{SemiEmpiModel}) 
for a source of intensity $\alpha$ after sifting procedure to the following expression:
\begin{equation}\label{Rsift}
R^\alpha_{\rm sift} = \frac{1}{2} f p_{\alpha} [Y_0 + 
\sum\limits_{\xi} p_{\xi} (1 - e^{-\alpha\eta_{\xi}(t)})]~,
\end{equation}
where $Y_0$ denotes a background signal from the four detectors, 
$\xi\in\{\rm H,V,D,A\}$ is a polarization type of the receiver channel.

\begin{figure*}[ht!]
\includegraphics[width=1\linewidth]{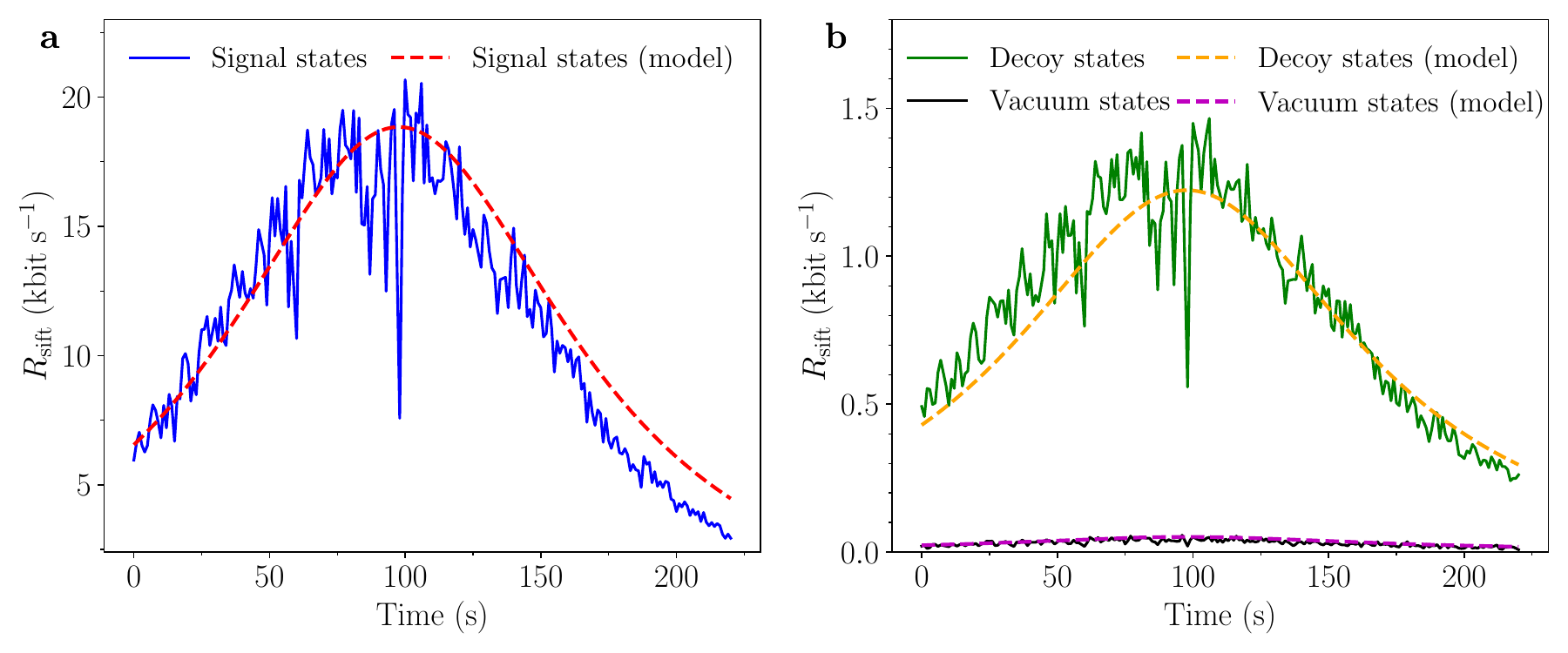}
\caption{The bit generation rate during the QKD experiment between the Micius satellite 
and the Zvenigorod ground station after the basis reconciliation. 
(\textbf{a}) The predicted and measured sifted key rate for a source of intensity $\mu=0.8$. 
(\textbf{b}) The predicted and measured sifted bit generation rate for a source of decoy states $\nu=0.1$ and vacuum states $\lambda = 0$. 
The atmospheric extinction coefficient in simulations is taken for clear weather conditions, that is, $\varkappa  = 0.22$.
\label{SiftKey}}
\end{figure*} 

The temporal filtering of quantum states reduces satellite and constant background noise by around five times. 
As a result, the background signal $Y_0$ per one sent pulse is equal to $\mathcal{N}/(5f)$ (for details, see Appendix~\ref{appendix:B}). 
WE note that it already contains afterpulse signals, owing to the special experimental way the data $\mathcal{N}$ is acquired.

Figure~\ref{SiftKey} depicts the obtained and the simulated sifted bit generation rates for 
the signal, decoy, and vacuum states over the satellite-to-ground QKD experiment under clear weather conditions ($\varkappa  = 0.22$). 
For the QKD experiment, the model accurately depicts the mean value of the bit generation rate. 
Although there is some difference between the simulated and experimental data, 
it might be explained by additional losses induced by pointing errors between the Micius satellite and the Zvenigorod ground station. 
During the quantum communication session, 
the resulting sifted key rate changes from around 20.7\,kbit/s at 600\,km to 2.9\,kbit/s at 1,100\,km, as shown in Fig.~\ref{SiftKey}a.

The QBER modeling has also been improved, 
taking into account the varying optical efficiency of the $\xi$-type channels and their intrinsic error $e_{\mathrm{det,}\xi}$ 
caused by measurement fidelity. 
The upper bound of an average intrinsic error for the $\xi$-type channel is considered as a time-dependent function comprised of two terms: 
\begin{equation}
\langle e_{\mathrm{det,}\xi} \rangle^\text{U} = e^{\rm Rx}_{\mathrm{det,}\xi} +  \langle e^{\rm Tx}\rangle~,
\label{edet}
\end{equation}
where $e^{\rm Rx}_{\mathrm{det,}\xi}$ is an intrinsic error for the $\xi$-type channel of the receiver 
when the input beam is perfectly polarized and $\langle e^{\rm Tx}\rangle$ 
is the average error probability over the four polarization states due to the real polarization contrast ratio of the transmitter optical scheme.

According to Ref.~\cite{Pan2017}, the average polarization contrast ratio of the Micius satellite source is 225:1, 
whereas the time-dependent part of the intrinsic errors $e^{\rm Rx}_{\mathrm{det,}\xi}$ 
is caused by the rotation of the polarization states and the birefringent elements in the receiver optical path. 
This may be measured empirically before the quantum communication session using the motorized half-wave plates.

Before the communication session, we compute the relative rotation of the polarization reference frames between the satellite and ground station based on the predicted trajectory of the Micius satellite~\cite{Han20, Bonato06}. 
Then the APS system of the receiver (see Fig.~\ref{scheme}) uses the calculated rotation angles to manage the motorized half-wave plate.
Therefore, to measure the intrinsic errors of the receiver, we simulate the rotation of each polarization state by sending classical 
laser beams into the ground station and then execute dynamical compensation following the computed rotation function~\cite{duplinsky}.
Figure~\ref{QBER}a illustrates the intrinsic errors $e^{\rm Rx}_{\mathrm{det,}\xi}$ of the Zvenigorod ground station 
after the dynamical compensation of the polarization state rotations for the Micius satellite passage on March 1, 2022.

Using Eq.~\ref{edet} the upper bound of the average number of errors in the sifted bits per second for a source of intensity $\alpha$ can be expressed as:
\begin{equation}
\langle N_{\rm err}\rangle^\text{U}_\alpha = \frac{1}{2} f p_{\alpha} [e_0 Y_0 + \frac{1}{4} \sum\limits_{\xi} \langle e_{\mathrm{det,}\xi}\rangle^\text{U} (1-e^{-\alpha\eta_{\xi}(t)})]~,
\label{QBER}
\end{equation}
where $e_0 = 1/2$ is the error rate of the background. 
Thus, the upper bound of the average QBER for a source of intensity $\alpha$ is given by:
\begin{equation}
\langle \mathrm{QBER}\rangle^\text{U}_\alpha = \langle N_{\rm err}\rangle^\text{U}_\alpha / R_{\rm sift}^\alpha~.
\end{equation}

\begin{figure*}[th!]
\includegraphics[width=1\linewidth]{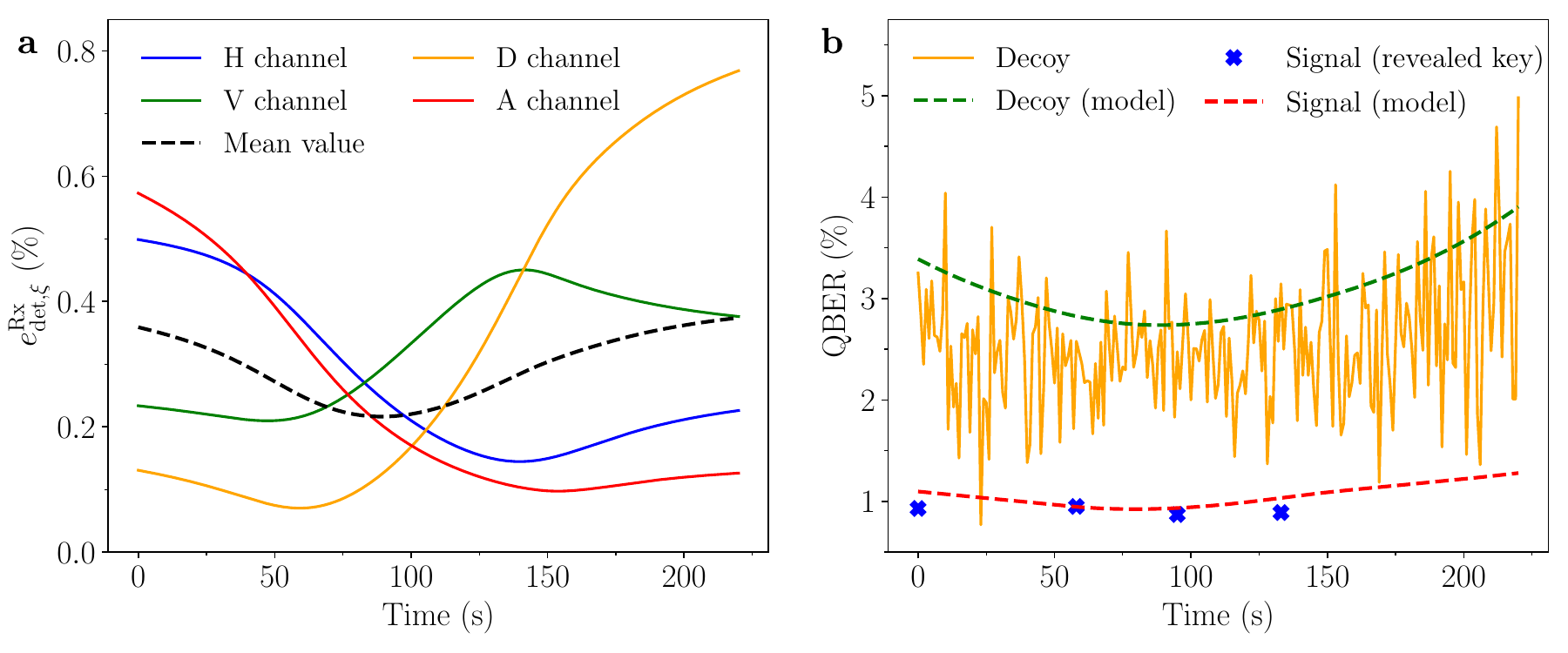}
\caption{Polarization characteristics and error analysis of the receiver.  
(\textbf{a}) The intrinsic error for each $\xi$-type channel of the Zvenigorod ground station 
when a test laser beam with a polarization contrast ratio of more than $10000:1$ is used. 
The function of polarization reference frame rotation corresponds to the Micius satellite passage on March 1, 2022. 
(\textbf{b}) The observed quantum bit error rate (QBER) and calculated upper bound of average one for 
a source of signal states and decoy states during the QKD experiment between the Micius satellite and the Zvenigorod ground station.
\label{QBER}}
\end{figure*}

The observed QBER in the satellite-to-ground QKD experiment and the predicted upper bound of average QBER 
for the signal and decoy states are shown in Fig.~\ref{QBER}b. 
In four time frames, the announced parts of the QBER for a source of signal states are 0.93\%, 0.95\%, 0.87\%, and 0.89\%, 
which correlate with calculated data. The experimental data on quantum errors for decoy states may be fully 
and securely opened according to the QKD decoy state protocol and therefore compared with the model predictions during 
the whole communication session. 
As a result, the model calculation accurately reproduces the shape of the average QBER. 
We note that the experimental and model QBER for vacuum states are close to 50\%, as expected.

Hence, the joint quantum communication experiment between the Micius satellite and the Zvenogorod ground station 
gives results that are quite close to the simulation outcomes according to a semi-empirical satellite-to-ground QKD model for realistic receivers. 
Thus, the successfully verified model is an excellent tool for simulating complicated and realistic QKD systems 
and analyzing their practical constraints and prospects.


\section{Security analysis and key rates}\label{sec4}
To reduce the possible knowledge of an eavesdropper, 
we estimate the secret key length  and use a random matrix for the shuffled key after the error correction step, 
with an inefficiency $f_{ec}= 1.44$~\cite{ErrCorr} 
(for details of the post-processing procedure, see Refs.~\cite{Kiktenko_2016,Fedorov2018,Kiktenko2020}. 
The decoy-state method is used to verify the lower bound on the single-photon counts. 
Therefore, if we consider statistical fluctuation and utilize Chernoff bounds with a failure probability 
\mbox{of $10^{-9}$~\cite{ SemiEmpiModel, Chernoff}}, the secure final key is 629,000 bits.

We can also consider higher-level security by taking into account the detection efficiency mismatch~\cite{Bochkov, Trushechkin, Ivchenko}. 
The problem is caused by the different channel optical efficiencies of the receiver and the complexity 
of designing detectors with the same quantum efficiency. 
Here we assume that the optical efficiencies of the \mbox{$\xi$-type} channels are equal 
and the detection efficiency imbalances are created by detectors with differing quantum efficiencies.

In our QKD experiment, we have four unbalanced detectors with the efficiencies $\eta_{z0}$, $\eta_{z1}$, $\eta_{x0}$ and $\eta_{x1}$, 
where the first subscript is a basis and the second one is a bit value. 
We assume that $\eta_{z1}\leq\eta_{z0}$ and $\eta_{x1}\leq\eta_{x0}$. 
Then the detection-efficiency mismatches in each basis are defined as $\eta_z=\eta_{z1}/\eta_{z0}$ and $\eta_x=\eta_{x1}/\eta_{x0}$. 
We take the values $\eta_z=0.60 $ and $\eta_x=0.51$. 
We note that we do not demand that the fractions of the efficiencies are precisely the values given above. 
The formulas for the secret key rate we used require that $\eta_{z1}/\eta_{z0}$ and $\eta_{x1}/\eta_{x0}$ are not smaller than 0.60 and 0.51,
respectively.
The ratio of the probability of signal reception in different bases is denoted as $t_{xz} \approx 1$, 
according to Fig.~\ref{Err&RawKey}a and Table~\ref{tab1}. 
We then assume that the beam splitter is set perfectly with the basis selection probability of $1/2$.

To perform a privacy amplification of the key, we use the calculations and results of Ref.~\cite{Bochkov, Ivchenko},
where the various methods of accounting for detection mismatch are addressed. 
Here we use an approach that allows us to obtain the maximum final key rate.

The formula for the final key generation rate for one impulse emitted by 
 Alice (transmitter)~\cite{Bochkov} is given by:
\begin{widetext}
\begin{eqnarray}
     K \geq p_z^2 p_{\rm det}^z \left[ h\left( \frac{1 - \delta_{z,z}}{2}\right) - h\left(\frac{1 - \sqrt{\delta_{z,x}^2 + \delta_{z,z}^2}}{2} \right) - f_{\rm ec} h\left(Q_z\right)\right]\qquad \nonumber \\ 
    +~ p_x^2 p_{\rm det}^x \left[ h\left( \frac{1 - \delta_{x,x}}{2}\right) - h\left(\frac{1 - \sqrt{\delta_{x,x}^2 + \delta_{x,z}^2}}{2} \right) - f_{\rm ec} h\left(Q_x\right)\right]~,
    \label{two}
\end{eqnarray}
\end{widetext}
\begin{equation}
    \delta_{z,z} = \frac{p_{z, 0} - p_{z, 1}}{p_{\rm det}^z}~, \qquad \delta_{z,x} = \frac{\sqrt{\eta_x}(t_z - 2q_x)}{p_{\rm det}^z}~,
\end{equation}
\begin{equation}
    \delta_{x,x} = \frac{p_{x, 0} - p_{x, 1}}{p_{\rm det}^x}~, \qquad \delta_{x,z} = \frac{\sqrt{\eta_z}(t_x - 2q_z)}{p_{\rm det}^x}~,
\end{equation}
where $p_x$ and $p_z$ are the probabilities of choice the $X$ and $Z$ bases by either Alice or Bob 
(for our case, $p_x = p_z = 1/2$), 
$p_{\rm det}^b$ denotes the probability of detecting the signal in the basis $b$, 
if Alice chose the same basis, $Q_x$ and $Q_z$ are the QBERs in signal states for appropriate bases, 
$f_{\rm ec} = 1.44$ is the error correction efficiency,
$p_{b, i}$ is the click probability of the detector with the bit value $i$, if Alice (transmitter) and Bob (receiver) chose the basis $b$, 
$t_b = p_{b, 0} + \frac{p_{b, 1}}{\eta_b}$ denotes the transparency of the channel for the $b$ basis, 
$q_b$ is a weighted mean erroneous detection rate in the $b$ basis (erroneous ones are taken with the weight $\eta_b^{-1}$ 
and erroneous zeroes are taken with the weight $1$).

This equation can be simply explained by physical considerations. 
First, we divide our data into two parts: one key is generated from the positions where Alice and Bob chose the $Z$ basis and another one -- 
where they chose the $X$ basis, and after that, these two keys are merged into one. 
Then, according to the Devetak and Winter theorem~\cite{devetak}, 
for each basis, the secret key rate is proportional to the difference between Eve's ignorance about Alice's string 
and Bob's ignorance about Alice's string with all information that legitimate users announce during communication. 
Bob's ignorance can be bound from above by binary entropy from quantum bit error rate; it is the last term in each bracket of Eq.~(\ref{two}). 
The first term (Eve's ignorance) contains some eavesdropper's information about the state. 
We can get rid of this using the entropic uncertainty relations and receive the first and second parts in each bracket of Eq.~(\ref{two}). 
Also, Eve can change the initial states $\rho_{AB}$ as she wants, 
but we can detect interference using conditions on the observable value of received bits and errors for different bases. 
The derivation is described in more detail in Appendix~\ref{appendix:C}.
This relationship is derived from general considerations about information and conditional entropy, 
and it is performed for our case of four detectors too. 
The main difference between the reasoning for two unbalanced detectors~\cite{Bochkov, Trushechkin} 
and our case is that both bases are used for key generation.

Hence, we estimate the final key rate using Eq.~(\ref{two}) applied to the divided data among two bases $X$ and $Z$.
As a result, 310,400\,bits of the secret key were finally shared between Alice and Bob. 

\section{Results and Discussion}\label{sec5}

\begin{figure}[th!]
\centerline{\includegraphics[width=1\linewidth]{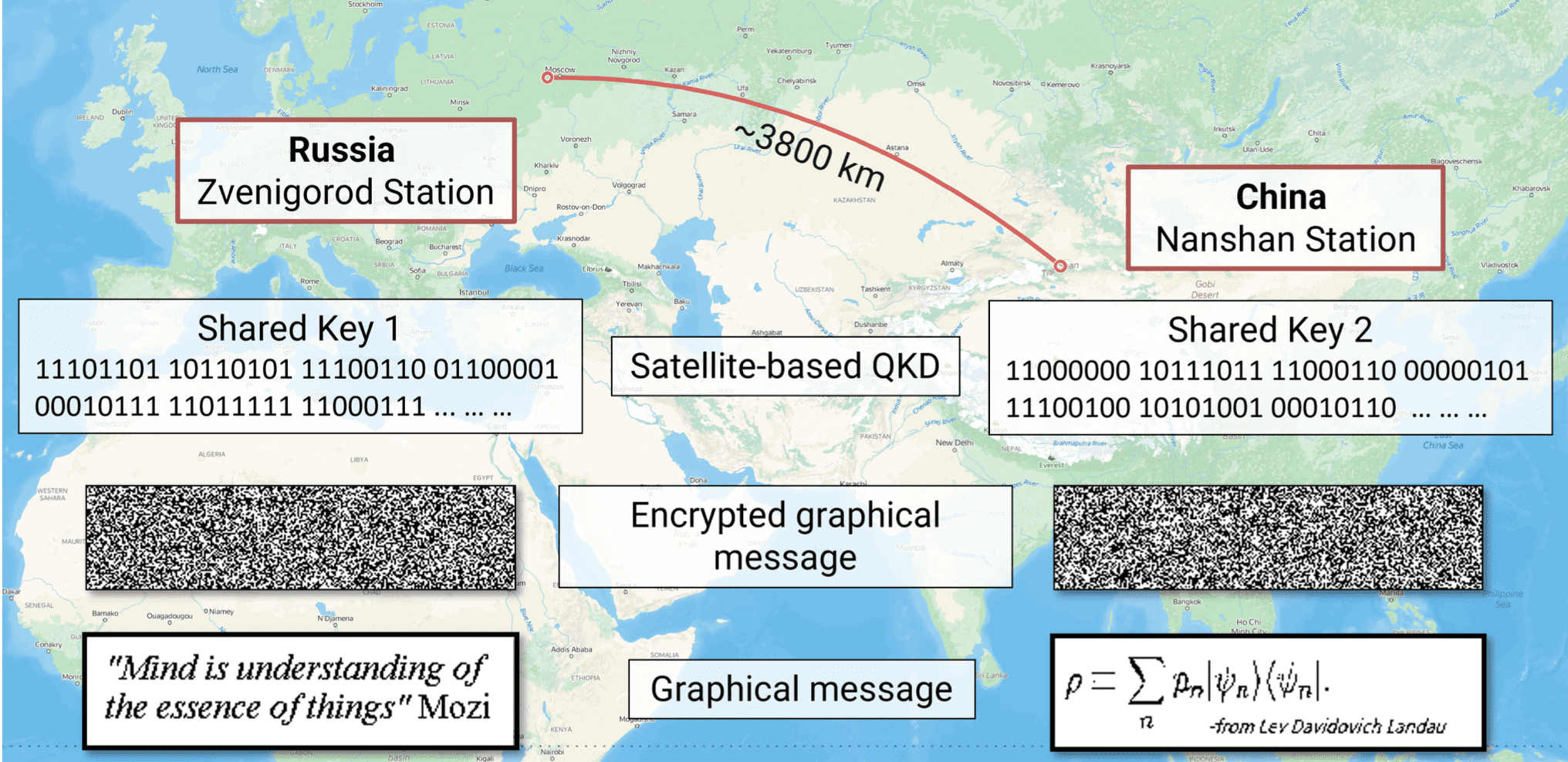}}
\caption{One-time-pad graphical message encryption and decryption between Zvenigorod (Russia) and Nanshan (China) ground stations. 
Each side used a basic bitwise XOR operation to encrypt/decrypt the image, with shared secret keys of 2 KB providing unconditional security.}
\label{Messages}
\end{figure}

Once we had achieved the shared secret key between Zvenigorod ground station and the Micius satellite, 
our colleagues from China carried out the same procedures with Nanshan ground station near Urumqi to get the shared key across these stations. 
As a result, a secret key is established between China and Russia at a distance of 3800 km on Earth.
Figure~\ref{Messages} illustrates the transfer of encoded messages with two shared keys totaling 4~kbytes. 
Each side used a basic bitwise XOR operation to encrypt and decrypt the image size of 2~kbytes, providing unconditional security.

We have reported satellite-to-ground QKD between the Micius satellite 
and the developed Zvenigorod ground station that enabled Eurasian-scaled secure message transfer. 
By simulating the parameters of the quantum communication link, 
we have validated our semi-empirical QKD model for the Zvenigorod ground station~\cite{SemiEmpiModel} 
with the results of the presented joint QKD experiment.  
Based on these experimental and simulated findings, we give a more secure analysis of the key rate than the general decoy state protocol, 
taking into account the detection efficiency mismatch. 
That kind of practical imperfection is common in realistic QKD devices, and we offer a possible solution to the problem in this research.

\section*{Acknowledgments}
We are grateful to our colleagues from the University of Science and Technology of China for their helpful assistance and numerous recommendations throughout the joint experiment with the Micius satellite.
The communications session has been completed on March 1, 2022. 
We thank A. Tayduganov and O. Fat’yanov for helpful discussions. 
We thank the Priority 2030 program at the National University of Science and Technology MISIS (Project No. K1-2022-027).

\appendix
\begin{widetext}
\section{\label{appendix:A} Supplementary Data}

\begin{table*}[h!]
\caption{Performance of Zvenigorod ground station \label{tabA1}}
\begin{ruledtabular}
\begin{tabular}{ccc}
	\textbf{}	& \textbf{Characteristic}	& \textbf{Data}  \\
	\colrule
	                  & Diameter ($D_T$)		     & 600~mm\\
	Telescope           & Focal length			       & 4800~mm\\
			            & Obstruction, ($\varepsilon$) & 0.73\\
\colrule
                        & Diameter	     & 70~mm\\
	Coarse control		& Focal length   & 350~mm\\
	(GT~\&~CAM2)	    & Field of view  & 12.4~mrad\,$\times$\,12.4~mrad \\
                        & Typical tracking error ($\sigma$) & $<\!100~\mu$rad\\
\colrule
                        & Wavelength   & 671~nm\\
   Beacon laser         & Power 	    & 3~W~@CW\\
			          & Divergence   & 3~mrad\\
\colrule
                        & Focal length (L1) 		       & 75~mm\\
			            & Field of view (CAM1)		       & 3.6~mrad\,$\times$\,3~mrad\\
Analysis and processing & Typical tracking error ($\sigma$)  & $<\!10~\mu$rad \\
    system (APS)        & Beam expander coefficient	       & 1\,:\,3\\
                        & Spectral filter (CWL)            & 850~nm \\
                        & Spectral filter (FWHM)           & 10~nm\\                                
\colrule
                            & Wavelength 			    & 850~nm\\
                            & Coupler focal length		& 18.5~mm\\
Polarization analyzer (PA)  & Polarizer transmittance   & 84.6\%\\
                            & Detector efficiency ($\eta_{\rm det}$) & $\thicksim\!60\%$\\
                            & Polarization contrast ratio (mean)        & $>\!350:1$\\
\colrule
                        & Spectral filter (CWL)         & 532~nm \\
Synchronisation         & Spectral filter (FWHM)        & 10~nm\\  
  module (SCM)		  & Coupler focal length		  & 18.5~mm\\
                        & Detector efficiency @532\,nm  & $\thicksim\!50\%$\\
\end{tabular}
\end{ruledtabular}
\end{table*}
\end{widetext}

\section{Noise estimation \label{appendix:B}}
The noise during satellite quantum communication is defined by the coefficients~$T$ and~$C$ in Eq.~(\ref{fit}). 
The coefficient~$T$ represents the mean number of sunlight photons at 850\,nm reflected from the Micius satellite surface to the ground station,
with the divergence angle equal to the quantum signal. 
Meanwhile, the background noise due to stray light and detector dark count is given by the coefficient~$C$. 

To determine the coefficient~$C$, we point the telescope at the night sky and record the count rate, which is $290\pm60$ clicks per second. 
However, the coefficient~$T$ cannot be measured in advance and must be estimated directly during quantum communication.
For instance, when the registered quantum signals are synchronized with the satellite clock, 
the noise may be found as the number of counts outside the time filtering frame and compared with the counting rate of registered vacuum states. 

\begin{figure}[h!]
\includegraphics[width=\linewidth]{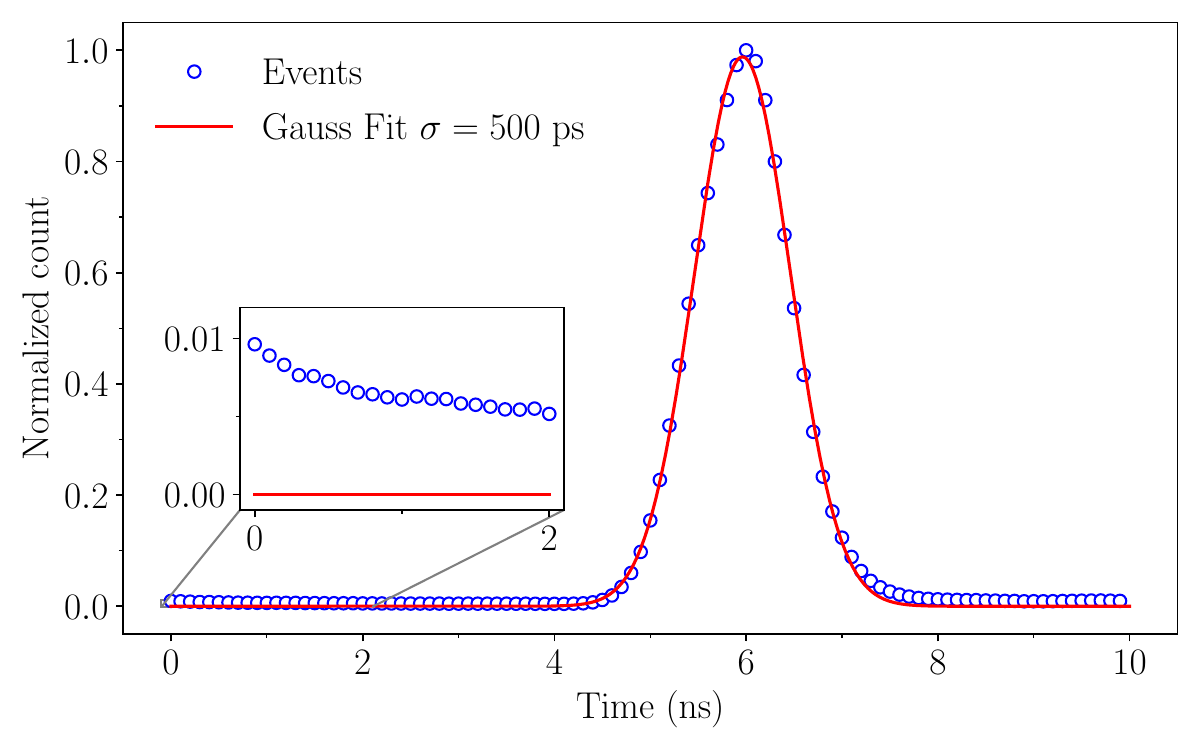}
\caption{The temporal distribution of 850 nm registered photons after the time synchronization over 220 seconds of communication session on 1 March 2022. Each time bin is 100\,ps. The counts are normalized, and Gaussian fitting yields a standard deviation of 500\,ps. \label{sync}}
\end{figure}

\begin{figure}[h!]
\includegraphics[width=\linewidth]{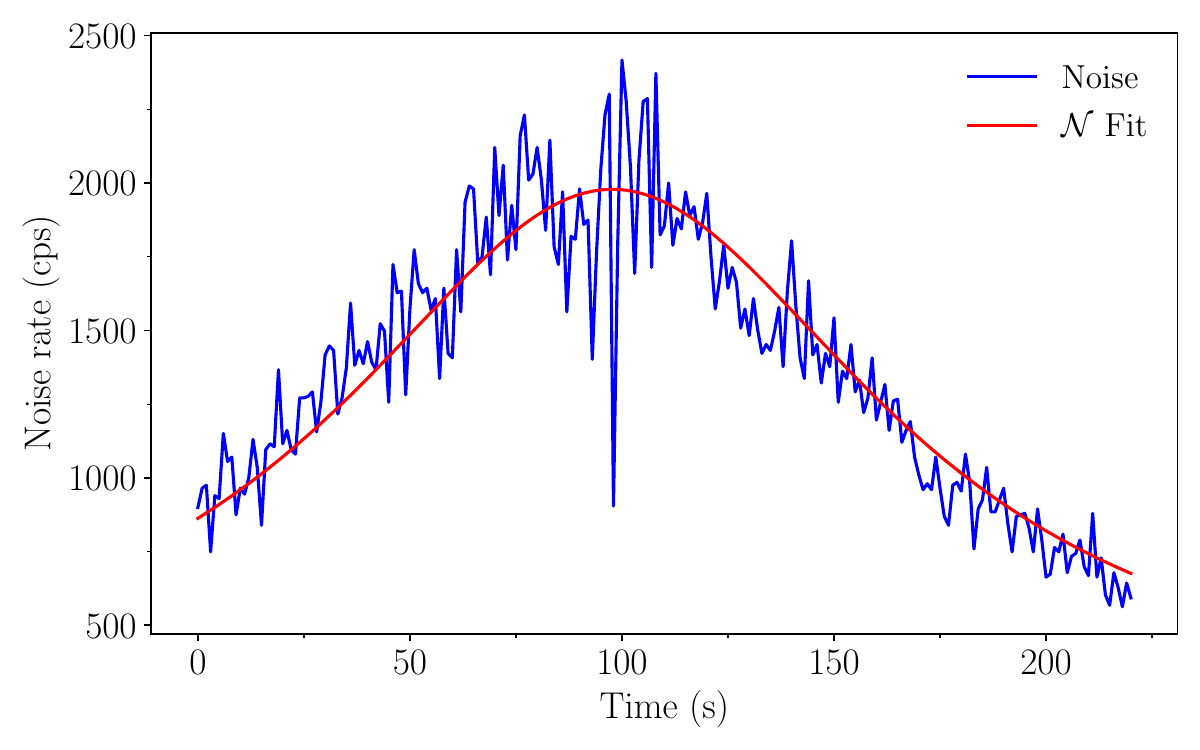}
\caption{ The noise rate for the satellite-to-ground QKD on March 1, 2022, obtained after clock synchronization, and the fitted function $\mathcal{N}$.
\label{noise} }
\end{figure}

The temporal distribution of quantum photons acquired after time synchronization is shown in Fig.~\ref{sync} 
with the distribution peak of counts corresponding to 6\,ns. 
To extract an overall noise, we compute the count rate in the time frame from $0$\,ns to $2$\,ns, which is opposite the distribution peak. 
Then the obtained data of the overall noise (see Fig.~\ref{noise}) are fitted with the following function: 
\begin{equation}
\mathcal{N}= T\eta(t) +C~,
\label{noiseeq}
\end{equation}
where $\eta (t)= \frac{1}{4}\sum \eta_\xi(t)$ denotes the time-dependent overall link efficiency of the communication channel 
[Eq.~(\ref{link})] with a fixed optical efficiency $\eta_{\rm opt}$ for the entire receiver of $27\%$ based on preliminary measurements.

As a result, the coefficient~$T$ is equal to \mbox{$1.8\cdot10^6\pm0.1\cdot10^5$\,photons per second}, 
and the atmospheric extinction coefficient $\varkappa$ is estimated at $0.25\pm0.05$. 
We note that to estimate the atmospheric extinction coefficient accurately, 
we should utilize a sufficiently high count rate, as it is done in Sec.~\ref{sec2.2} for the total count rate of the raw key.

\section{Derivation of the secret key rate \label{appendix:C}}

A complete analysis for the case of detection-efficiency mismatch is presented in Refs.~\cite{Bochkov, Trushechkin}. 
Here we use these articles as a guide and present the general ideas on how to adopt this theory in the case of four detectors. 
The main result of this section is Eq.~(\ref{two}) for the secret key rate.

Denote $\rho$ is the density operator distributed by Eve (according to the equivalent entanglement-based formulation, 
also known as the source replacement scheme). 
Then, we express the process of imperfect measurement as two stages: 
the attenuation on detectors $\rho\mapsto G_b\rho G_b$, 
where $G_b$ is the attenuation operator corresponding to the measurement in the basis $b$, followed by a perfect measurement. 
The operator $G$ is not unitary but turns out to be Hermitian. 
$\Tr G_b\rho G_b=p_{\rm det}^b$ is the detection probability in the basis $b\in{X,Z}$ and $\widetilde\rho=G_b\rho G_b/p_{\rm det}^b$.

According to Devetak and Winter theorem from Ref.~\cite{devetak}, the secret key rate is expressed as follows:
\begin{eqnarray}
    K = p_{x}^2 p_{\rm det}^x\left[ H(X|E)_{\widetilde{\rho}} - H(X|B)_{\widetilde{\rho}} \right]+\quad \nonumber  \\ 
    +~p_{z}^2 p_{\rm det}^z\left[ H(Z|E)_{\widetilde{\rho}} - H(Z|B)_{\widetilde{\rho}} \right]~,
    \label{A1}
\end{eqnarray}
where $p_x = p_z = 1/2$.

The reasoning in both bases is identical and is transformed by interchanging $X$ and $Z$. 
So we will give reasoning for only the first line in Eq.~(\ref{A1}). 
The first term contains the Eve subsystem. To get rid of this, according to Ref.~\cite{entropy2010}, 
we can approximate this term by using the entropic uncertainty relations:
\begin{equation}
    H(X|E)_{\widetilde{\rho}} + H(Z|B)_{\widetilde{\rho}} \geq 1~.
    \label{A2}
\end{equation}

The second term in brackets can be bounded from above by binary entropy from QBER for the signal states; 
one can say that this is Bob's ignorance about Alice's string. 
The factor $f_{\rm ec}$ appears due to imperfections in the error correction procedure. 
Substituting the uncertainty relation Eq.~(\ref{A2}) into Eq.~(\ref{A1}) for the secret key rate, 
we receive the next lower bound for the generation rate:
\begin{eqnarray}
    K \geq p_{x}^2 p_{\rm det}^x\left[ 1 - \min_{\rho_{AB} \in S_z} H(Z|B)_{\widetilde{\rho}} - f_{\rm ec} h(Q_x)\right]+~~~\nonumber \\ 
    +~p_{z}^2 p_{\rm det}^z\left[ 1 - \min_{\rho_{AB} \in S_x} H(X|B)_{\widetilde{\rho}} - f_{\rm ec} h(Q_z)\right]~,~~
    \label{A3}
\end{eqnarray}
where $\rho_{AB} = \Tr_E\rho$ is the density operator of Alice’s and Bob’s subsystem. 
The imposed constraints on the initial density matrix that define the spaces $S_x$ and $S_z$ are taken from Ref.~\cite{Trushechkin}.

According to Ref.~\cite{Trushechkin}, we estimate the conditional entropy for the $Z$ basis as follows:
\begin{eqnarray}
    \min_{\rho_{AB} \in S_z} H(X|B)_{\widetilde{\rho}} \leq 1 - h\left( \frac{1 - \delta_{z,z}}{2}\right) - \qquad\nonumber \\ 
    -~h\left(\frac{1 - \sqrt{\delta_{z,x}^2 + \delta_{z,z}^2}}{2} \right)~,
    \label{A4}
\end{eqnarray}
where 
\begin{equation}
    \delta_{z,z} = \frac{p_{z, 0} - p_{z, 1}}{p_{\rm det}^z}~,  \qquad \delta_{z,x} = \frac{\sqrt{\eta_z}(t_z - 2q_x)}{p_{\rm det}^z}~.
\end{equation}

The upper bound on the entropy for the $X$ basis is made similar to Eq.~(\ref{A4}). 
Thus, we obtain the desired Eq.~(\ref{two}) by substituting Eq.~(\ref{A4}) and the corresponding one for the $X$ basis in Eq.~(\ref{A3}).

\bibliography{bibliography.bib}

\end{document}